# Leaping shampoo and the stable Kaye effect

Michel Versluis*, Cor Blom, Devaraj van der Meer, Ko van der Weele, and Detlef Lohse

*Physics of Fluids, Faculty of Science and Technology*

*University of Twente, P.O. Box 217, 7500 AE Enschede, The Netherlands.*

\* *corresponding author:*

    Dr. Michel Versluis
    Physics of Fluids
    Science and Technology
    University of Twente
    P.O. Box 217
    7500 AE Enschede
    The Netherlands

    E-mail:    `m.versluis@utwente.nl`
    Phone:    +31 53 489 6824
    Fax :    +31 53 489 8068
    website:    `http://pof.tnw.utwente.nl/`

## Abstract

Shear-thinning fluids exhibit surprisingly rich behaviour[1]. One example is the Kaye effect[2] which occurs when a thin stream of a solution of polyisobutylene in Decalin is poured into a dish of the fluid. As pouring proceeds, a small stream of liquid occasionally leaps upward from the heap. This surprising effect, which lasts only a second or so, is named after its first observer A. Kaye, who could offer no explanation for this behaviour. Later, Collyer and Fischer[3] suggested from 250 frames per second cine recordings that the fluid must be highly shear thinning as well as elastic and 'pituitous'[*]. In addition, they concluded that a rigid surface is required to back the reflected liquid stream. While the words *bouncing* and *reflection* are associated with non-continuous and elastic effects, we will show here that the Kaye effect is in fact a continuous flow phenomenon. We show that the Kaye effect works for many common fluids, including shampoos and liquid soaps. We reveal its physical mechanism (formation, stability and disruption) through high-speed imaging. The measurements are interpreted with a simple theoretical model including only the shear thinning behaviour of the liquid; elastic properties of the liquid play no role. We show that the Kaye effect can be stable and that it can be directed. We even demonstrate a stable Kaye effect on a thin soap film excluding the necessity of a rigid backing surface.

---

[*] slimy or sticky

A wide variety of materials show shear-thinning behaviour[1], i.e. the viscosity of the fluid decreases with increasing shear rate. Polymer melts and polymer solutions[4], emulsions and dispersions[5] are classes of materials which often display such a non-Newtonian behaviour. Examples include biological fluids like blood and saliva and so-called synovial fluids found e.g. in the knee joint, food-engineered products like ketchup and yoghurt, cosmetic creams, gels and liquid detergents from the personal care industry and drilling mud, cement pastes and latex paints from the building and construction industry. All these fluids have in common that they are stable at rest, yet in their specific application their resistance to flow must be low. Shear thinning is related to reversible structural break-down of the materials due to flow-induced stress.

An extreme example of shear-thinning behaviour is the Kaye effect[2], shown in Figure 1 (see Supplementary Information, Video S1–S4). Here, a shear-thinning shampoo is poured from a height of approximately 20 cm in a thin stream of thickness 0.4 mm and visualized using a high-speed camera at 1000 frames per second. At first, the fluid will curl and wrinkle as any highly viscous fluids such as honey, syrup (see Supplementary Information, Video S5) or silicone oil would do[6,7] forming a viscous heap (Fig. 1.1). At some instant, due to a favourable geometry, the incoming jet will slip away from the heap. While for a viscous Newtonian fluid such a slip would only lead to a small disturbance in the wrinkling or coiling pattern, in the shear-thinning shampoo the resulting high shear rate forms a low viscosity interface leading to an expelled jet at low inclination (Fig. 1.2). Meanwhile the incoming jet will exert a vertical force on the viscous heap forming a dimple. The dimple deepens because of the sustained force exerted onto it by the incoming jet thereby erecting the outward going jet (Figs. 1.3-1.5). The inclination of the streamer steepens until it hits the incoming jet (Fig. 1.6) and disturbs or even interrupts the in-flow, thereby halting the Kaye effect. As will be shown later, during the jetting stage there is no net mass transport to the viscous heap and therefore the heap will slowly decay at a timescale of seconds. As soon as the jetting stops, the fluid will form a new heap from which a successive slip event can occur, repeating the series of events.

A snapshot of a streamer ejecting from the viscous heap is shown in Fig. 1A. The ejected streamer generally has the shape of a lasso, with one end fixed to the viscous heap and the other end directly connected to the incoming jet. As the in-flow proceeds the lasso extends, leaving a stagnant branch and a flowing branch of fluid (see Fig. 1B and Figs. 1.3 and 1.4). The flowing branch shoots forward and remains flowing. One striking phenomenon that is observed in all experiments is that the outgoing jet diameter is considerably larger than the incoming jet diameter, see Fig. 1B, indicating a lower velocity of the leaping jet. The velocities of the incoming and outgoing jet, $V_{in}$ and $V_{out}$, were measured by tracking microbubbles contained in the transparent fluid (see e.g. Supplementary Information, Video S4 andS6). The diameter of the streamer $2R_{out}$ is found to be thicker to that of the incoming jet $2R_{in}$, quantitatively following the continuity equation $V_{in}R^2_{in}=V_{out}R^2_{out}$. The lower velocity relates to viscous dissipation in the dimple structure, as will be shown later. The velocity of the outgoing streamer as it leaves the heap was determined independently by measuring the profile of its perfect parabolic trajectory and these values were found to be in very good agreement.

Upon impact the incoming jet exerts a force on the underlying viscous heap. From the vertical component of the velocities the momentum transfer was calculated. In our range of experiments the force was in the order of $5\times10^{-3}$ N. The vertical force results in a spoon-like dimple structure, which facilitates the jet to leap forward, see Fig. 2A and Supplementary Information, Video S7. High-speed imaging, as was used to capture this image, allows for a detailed study of this fast event, yet the dimple structure is also visible to the unaided eye. In Collyer and Fischer's work[3] the steepening trajectory of the outgoing streamer was explained by a collapse of the heap which flattens, thereby presenting a smaller angle of incidence to the incident stream and causing the 'bouncing' streamer to steepen its trajectory. However, the current explanation of a continuous flow mechanism and the formation of a dimple that deepens with time discards this picture of a bouncing streamer.

We investigated several conditions for the release height, width and velocity of the incoming jet for the occurrence of the Kaye effect. From these experiments it followed that for small release heights, where the incoming vertical speed of the liquid is relatively low, the liquid wrinkles like any viscous fluid would do. At higher incoming velocities (higher release heights) the liquid stream slides away to the side due to its shear-thinning property, however the expelled jet is rapidly brought to a standstill due to viscous dissipation at the underlying surface. Then, by a small increase in the release height, the expelled streamer will lead to the formation of a jet. Such a critical release height was consistently found for all studied mass flows. For even higher release heights, in the presence of a jet, we always observed a large difference between leap height and release height, reflecting that a considerable amount of energy must be dissipated in the dimple.

The dissipation mechanism in the dimple structure was captured in a simple energy balance model. Our model assumes a thin shear layer between the flowing fluid and the viscous heap. At the high shear rates present in the shear layer the viscosity of the shear-thinning fluid drops by several orders of magnitude. Nonetheless, some dissipation occurs through viscous friction. The model described here calculates the incoming power and the dissipation rate in the dimple structure. Then from the difference, the velocity and leap height can be obtained. Figure 2B shows the relevant parameters for this model. The power $P_{in}$ of the incoming stream of fluid is determined by its mass flow $\dot{m}$ and its velocity $V_{in}$ when reaching the viscous heap, $P_{in} = \frac{1}{2}\dot{m}V_{in}^2$. At the point of impact the radius $R_{in}$ of the stream equals $R_{in} = [\dot{m}/(\pi \rho V_{in})]^{1/2}$, with $\rho$ the density of the fluid. Viscous friction between the flowing fluid and the stagnant heap in the dimple structure (contact length $L$) leads to energy dissipation, hence to a loss in flow velocity. In the thin shear layer of thickness $\delta$ between the flowing fluid (velocity $V$) and the stagnant heap (velocity zero) we assume a linear velocity profile. This leads to a constant shear rate $\dot{\gamma} = V/\delta$, which in our experiments is of the order of $10^4$ s$^{-1}$. The shear stress $\tau$ in the layer is then given by $\tau = \eta \dot{\gamma}$, where $\eta$ is the shear rate-dependent viscosity.

The viscosities of the shear-thinning fluids used in this study were measured independently with a coaxial cylindrical rheometer, see Fig. 3, for shear rates up to $10^3$ s$^{-1}$, which was the experimental upper limit due to air entrainment and foam formation. In the thin shear layer of the leaping jet ($V$ = 1m/s, $\delta$ = 100μm) the shear rates are even higher. Therefore we extrapolate the experimental results following the established empirical equation given by Cross[8]:

$$\eta(\dot{\gamma}) = \eta_\infty + \frac{\eta_0 - \eta_\infty}{1+(\dot{\gamma}/\dot{\gamma}_c)^n} \quad , \tag{1}$$

where $\eta_0$ and $\eta_\infty$ represent the zero-shear-rate and infinite-shear-rate viscosity respectively, and where the critical shear rate is represented by $\dot{\gamma}_c$. Typical values for the shear-thinning liquids used in this study were $\eta_0$ = 5-10 Pa.s and $\dot{\gamma}_c$ = 10-20 s$^{-1}$; the exponent $n$ approaches unity for our studied liquids. The viscosity at infinite shear rate was taken that of water, $\eta_\infty$ = 1 mPa.s. The reason for this is that the long soap molecule structures that build up viscosity at rest break up under shear and in the limit of high shear rates become so small that we basically end up with a solution containing isolated soap molecules. For the shear rates in our experiment we find $\dot{\gamma} \gg \dot{\gamma}_c$ and Eq. (1) can in very good approximation be written as

$$\eta(\dot{\gamma}) \approx \eta_\infty + \frac{\eta_0 \dot{\gamma}_c}{\dot{\gamma}} = \eta_\infty + \frac{\eta_0 \dot{\gamma}_c \delta}{V} \quad . \tag{2}$$

The dissipated power $dP_{diss} = V\, dF$ over a small contact length $dx$ measured along the dimple structure is calculated from the shear force $dF = \tau\, dA$ acting on the contact area $dA = \pi R\, dx$ and should be balanced by the decrease of the kinetic energy passing per second $d\dot{E}_{kin} = \tfrac{1}{2}\dot{m}\left[(V(x+dx))^2 - (V(x))^2\right]$. This eventually leads to the differential equation

$$\dot{m} V\, dV = -V\left(\eta_\infty + \frac{\eta_0 \dot{\gamma}_c \delta}{V}\right)\frac{V}{\delta}\pi\sqrt{\frac{m}{\pi \rho V}}\, dx \quad , \tag{3}$$

which can be non-dimensionalized by introducing the velocity- and length-scales $\beta$ and $\Delta$ defined as $\beta \equiv \eta_0 \dot{\gamma}_c \delta / \eta_\infty$ and $\Delta^2 \equiv [\rho \dot{m} \eta_0 \dot{\gamma}_c / \pi](\delta/\eta_\infty)^3$, respectively. With $U = V/\beta$ and $\xi = x/\Delta$ Eq. (3) becomes

$$\frac{dU}{d\xi} = -\left[\sqrt{U} + \frac{1}{\sqrt{U}}\right], \tag{4}$$

which needs to be integrated along the dimple structure, i.e., with $x$ ranging from 0 to $L$, and $V$ from $V_{in}$ to $V_{out}$. The exact result of this integration gives an implicit equation for $U_{out}(U_{in})$:

$$\left[\sqrt{U_{out}} - \arctan(\sqrt{U_{out}})\right] = \left[\sqrt{U_{in}} - \arctan(\sqrt{U_{in}})\right] - \tfrac{1}{2}L/\Delta . \tag{5}$$

from which the dependence of the leap height as a function of the release height directly follows, see Fig. 4A (lines). The model reflects the experimentally observed threshold behaviour, which can immediately be understood by using the approximation $y - \arctan y \approx y^3/3$, valid for small $y$ (small $V$ in relation to $\beta$); Then one obtains $U_{out}^{3/2} \approx U_{in}^{3/2} - \tfrac{3}{2}L/\Delta$, i.e. there will be no outward velocity below a certain minimal $U_{in,M} = \left(\tfrac{3}{2}L/\Delta\right)^{2/3}$ which is found by taking $U_{out} = 0$ in Eq. (5). Physically this reflects that the Kaye effect can only be stable if the kinetic energy of the incoming stream exceeds the total dissipation in a dimple of length $L$. In dimensional quantities the threshold velocity reads in this approximation:

$$V_{in,M} \approx \left(\frac{\pi}{\rho \dot{m}}\right)^{1/3} \left(\tfrac{3}{2}\eta_0 \dot{\gamma}_c L\right)^{2/3} . \tag{6}$$

Most importantly, this implies that in leading order the minimum velocity does *not* depend on the thickness of the shear layer $\delta$, the only parameter in our model which cannot be well estimated from our experiments.

The occurrence of the Kaye effect was studied experimentally by varying the release height between 5 cm and 30 cm for a mass flow of 0.05 g/s and 0.2 g/s. Figure 4A shows the measured leap height vs. release height (circles). It is seen that the experimental results compare well with the model predictions. The model contains two free fit parameters: the shear layer thickness $\delta$ and the contact length within the dimple structure $L$. In the examples shown here $\delta$ was kept at 100 μm, while $L$ was taken 3.8 mm and 5.5 mm for a mass flow of 0.05 g/s and 0.2 g/s, respectively. A larger mass flow indeed leads to a larger dimple structure, as observed in experiment. Fig. 4B illustrates the corresponding power of the incoming jet, the dissipation rate and the power of the outgoing jet calculated as a function of the release height. For high release heights the incoming power exceeds the dissipation rate, leading to a successful leap. By lowering the release height the power of the outgoing jet decreases and at a certain point becomes zero. This defines the critical release height in the model; below this point no leaping occurs.

One observation for the Kaye effect is that, as it develops (following e.g. Fig. 1.1-1.6), the dimple in the viscous heap deepens. Therefore the contact length increases from its initial optimum contact length to larger and larger values, thereby increasing the dissipation. The increase in dissipation rate can also be modelled and continues until it reaches a critical value equal to the incoming power, thereby halting the Kaye effect. This mechanism is observed experimentally in several occasions. The principal disruptive mechanism to end the Kaye effect, however, is that of the physical interaction of the outgoing and the incoming jets. This however can be prevented altogether, as discussed in the following.

The conventional setup for the Kaye effect has the disadvantage that, because of its geometry, the erected outgoing jet interferes with the incoming jet leading to a disruption of the Kaye effect (see Fig. 1.6). By tilting the bottom surface we succeeded in constructing a system displaying a *stable* Kaye effect which lasts for many minutes. A photograph of this system is shown in figure 5. The surface is covered with a thin film of the same material flowing from a small reservoir on top. The starting principle of the stable Kaye effect is similar to that of the fluid falling on the horizontal plate. The outgoing jet may originally eject in any arbitrary direction, however, the jet will always orient itself in the direction downward along the surface, as this will minimize the contact length and therefore the dissipation. The expelled jet lands on the coated surface and as the jet steepens the inclination angle increases, leading to a secondary Kaye effect. This process can repeat itself several times along the inclined plane and we have been able to create a *cascading* Kaye effect with up to seven leaps. The cascading effect is also very stable. It is the balance of the incoming power and the cumulative viscous dissipation within each dimple which eventually terminates the cascade after a finite number of steps.

We investigated the necessity of several conditions for the Kaye effect to occur, starting with the presence of a backing plane. We found that a rigid backplane is not required for a support of the Kaye effect. This was demonstrated by pouring a thin stream of shear-thinning fluid on top of a deep container filled with the same material. We found no difference with a Kaye effect performed on a flat backing plane. We even tried and successfully created a Kaye effect on a thin liquid soap film, as demonstrated in figure 6A. The soap film was stretched across a thin metal wire of rectangular shape, the film thickness being approximately 50 μm. Shear thinning liquid soap was poured from a height of 20 cm and interestingly, also in this case the fluid leaps. When the plane of the film was inclined the Kaye effect was stabilized (as in the case of the cascade) and could be directed. The effect continues for several minutes, or as long as the soap film lasts through film drainage. As this setup is optically thin it allows for a detailed study of the jet bending structure, see Fig. 6B and Supplementary Information, Video S8. First, the jet curvature and its thickness (directly related to its velocity) can be deduced. Secondly, the forces exerted by the supporting structure can be studied through our understanding of thin film curvature, surface tension and corresponding normal forces. Finally, not studied in detail at present, normal stresses of the rheological material may also account in part for the support of the bending of the incoming jet. In any case, these experiments disprove the claim of Collyer and Fischer[3] that a rigid backing plane is required for the Kaye effect.

Conservation of mass was demonstrated before based on the velocities and radii of the incoming and outgoing streamers. Continuity was also demonstrated in an experiment with two similar shear-thinning fluids, one being transparent, the other being white opaque. The transparent fluid functioned as the supporting base substance while the opaque one was poured on top of it. From these experiments it followed that no mixing occurred and that the two fluids in fact remain separated. This was also demonstrated in another way by using the fluid streamers as laser light guides. In a setup with a transparent fluid laser light from a Helium-Neon laser beam was introduced through the nozzle opening. The vertical stream then acts as a liquid laser light guide, as the total internal reflection at the free interface keeps the laser light contained in the fluid. While the stream ends up in a viscous heap the laser light will escape from it as the angle is in many cases smaller than the critical angle. It will therefore brightly illuminate the heap, see Fig. 7A. During the occurrence of the Kaye effect we observe that the laser light remains within the outgoing streamer (Fig. 7B) (see Supplementary Information, Video S9). The light now escapes at the position where the outgoing streamer ends, in the cascading Kaye effect this can be after five consecutive leaps or more (see Supplementary Information, Video S10). The fact that the laser light does not penetrate to the viscous heap indicates the structural changes of the fluid through shear in the thin shear layer. The laser detection method therefore shows that the flowing stream is in fact decoupled from the viscous heap, both in an optical sense as well as in a fluid mechanical sense.

In conclusion, and in contrast to what has previously been reported, the Kaye effect is comprised of the efficient bending of a low-viscosity liquid jet by a favourable spoon-like dimple formed in a viscous heap of fluid. For a highly shear-thinning fluid, such as those used in our study, these two conditions can develop within the same liquid.

**Supplementary information** includes digital high-speed video recordings. Two additional video files show an example of the Kaye effect taken with a regular video camera.

**Acknowledgements** The experimental assistance of Marc Harleman and Arjan van der Bos is gratefully acknowledged. The authors also wish to thank Gert-Wim Bruggert for his technical assistance.

**Author information**. The authors declare no competing financial interests. Correspondence and request for materials should be addressed to M.V. (`m.versluis@utwente.nl`).

**Figure legends**

**Figure 1.** The Kaye effect for leaping shampoo. Top row: Six snapshots of a shear-thinning fluid (shampoo) showing the Kaye effect[2]. The images were recorded with a digital high-speed camera operating at 1000 frames per second. (1) Formation of a viscous heap through piling, buckling and coiling at t = 0 ms. (2) Ejection of a thin streamer of fluid initiates the Kaye effect at t = 30 ms. (3-5) The jet rises at t = 70 ms, 160 ms, 280 ms. (6) The outgoing jet disrupts the incoming jet which leads to its collapse and the end of the Kaye effect at t = 380 ms. Note that the viscous heap decays at a relatively slow timescale (order 1 second). See Supplementary Information, Video S3. **A.** Inception of the Kaye effect. A streamer slips away from the viscous heap. One part of the lasso is connected to the viscous heap, therefore it cannot flow and it will drop onto the bottom surface (see **B**). The other part of the lasso is directly connected and fed by the incoming streamer and it will therefore flow ahead. See Supplementary Information, Video S4.

**Figure 2.** Dimple structure in the viscous heap. **A.** High-speed recording showing the dimple structure (exposure time 150 µs). It is seen here that the flow is continuous. Note the difference in radii of the incoming and outgoing streamer. The viscous heap and stagnant part of the initial lasso have meanwhile decayed. See Supplementary Information, Video S6 and S7. **B.** Schematic view of a cross section of the flow within the dimple structure used in the viscous dissipation model. $V$ indicates the local velocity of the jet. $L$ represents the interaction length of the flowing jet with the stagnant heap. The thickness of the shear layer is indicated by $\delta$ and $\eta$ represents the shear rate dependent viscosity.

**Figure 3**. Viscosity versus shear rate for a typical fluid used in our experiments measured in a coaxial cylindrical rheometer (dots). The graph clearly shows the shear thinning properties of the fluid. The well-known problem of foaming prevents measurements above a shear rate of $10^3$ s$^{-1}$, but the high shear rate behaviour can be extrapolated using the established empirical equation given by Cross[8] (solid line). This particular graph was fitted with the following parameters: a zero-shear-rate viscosity $\eta_0$ = 6.0 Pa.s, an infinite-shear-rate viscosity of $\eta_\infty$ = 1 mPa.s (fixed to the viscosity of water), a critical shear rate $\dot{\gamma}_c$ = 18 s$^{-1}$ and an exponent of $n$ = 1 (thin grey line indicates a slope of -1).

**Figure 4. A.** Experimental leap height versus release height for a mass flow of 0.05 g/s (red filled circles) and 0.2 g/s (blue open circles). The lines represent the modelled leap height as a function of the release height using the parameters given in B. **B.** Power diagram for the Kaye effect. The incoming, dissipated and outgoing (leaping) power plotted as a function of the release height for a mass flow of 0.05 g/s (red solid line) and 0.2 g/s (blue dotted line). At the critical height, indicated by the thin vertical lines, the incoming power equals the dissipated power. Calculations are based on a shear layer thickness of 100 μm and a contact length of 3.8 mm and 5.5 mm for a mass flow of 0.05 g/s and 0.2 g/s, respectively.

**Figure 5.** Cascading and stable Kaye effect. This phenomenon occurs on an inclined plane prepared with a thin film of the same material. The fluid was found to leap at up to 7 times. In addition, this effect was found to be stable and lasted up to ten minutes which was as long as the reservoir was supplying a stable incoming stream of fluid.

**Figure 6.** Stable Kaye effect on a liquid soap film. **A.** The stable Kaye effect demonstrated for a thin jet of a shear-thinning fluid (liquid hand soap) leaping off a 50 μm thin film of the same material. The film is locally deformed by the forces exerted onto it by the bending jet, however the surface tension is strong enough to keep the film intact. **B.** As the film is optically thin this stable setup reveals the important physical parameters and mechanisms for the Kaye effect. See Supplementary Information, Video S8.

**Figure 7.** Laser light guide experiment for the Kaye effect. Laser light from a Helium-Neon laser was introduced into a transparent shear thinning fluid through the nozzle opening of the reservoir. **A.** The laser light illuminates the viscous heap as the streamer terminates inside the heap. **B.** For the Kaye effect the light is transported away from the heap within the outgoing streamer indicating the continuity of the jet. See Supplementary Information, Video S9 and S10.

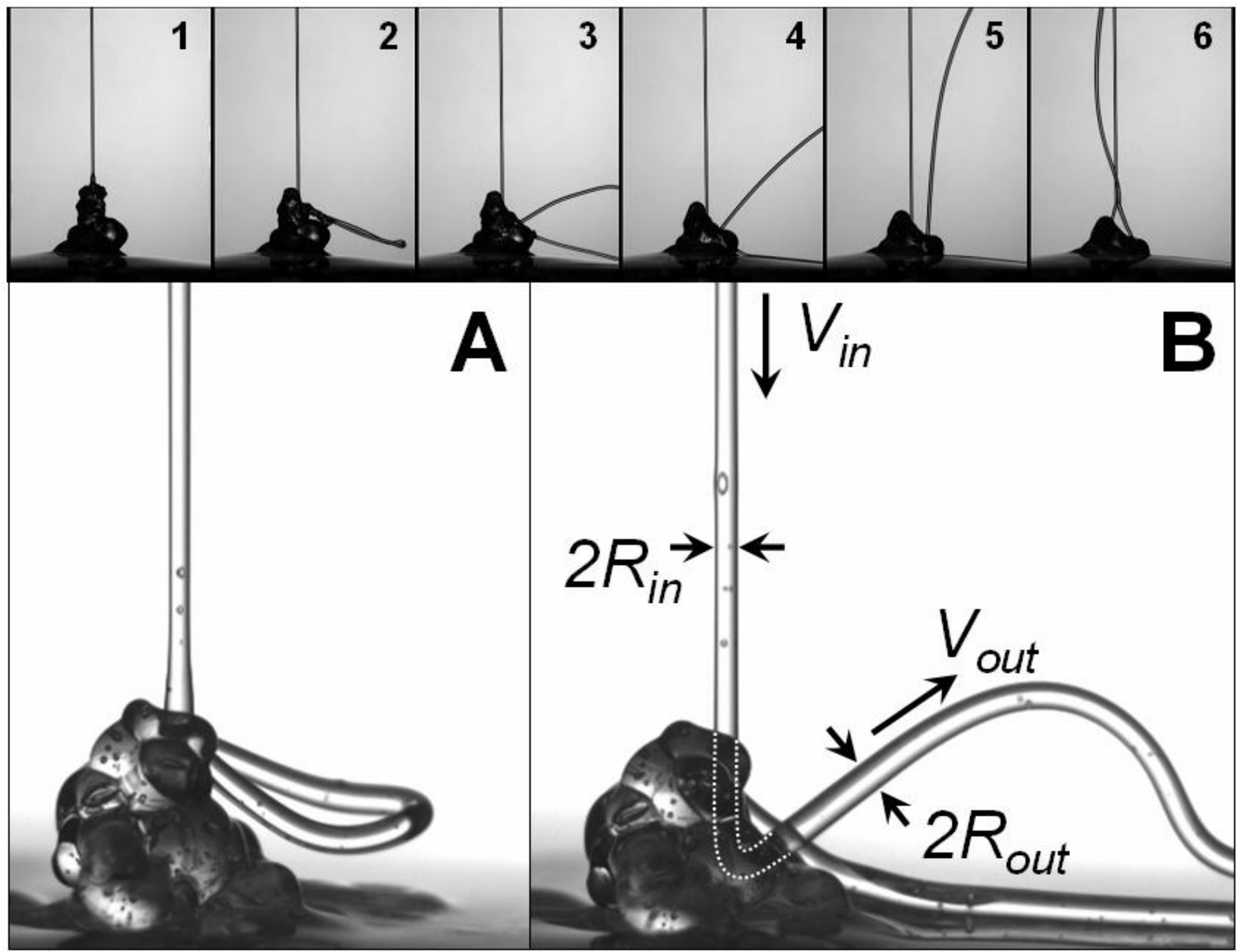

Fig. 1

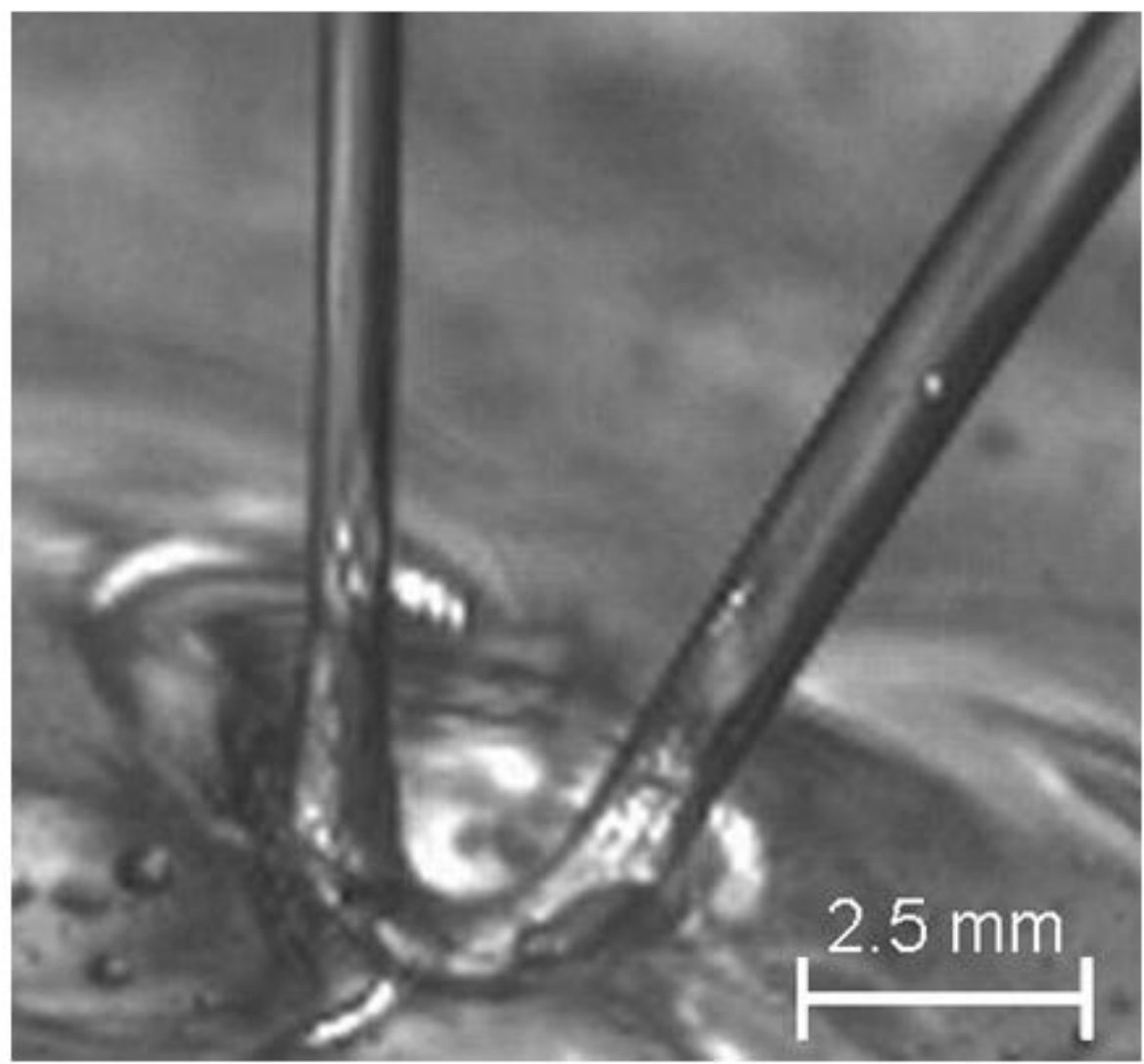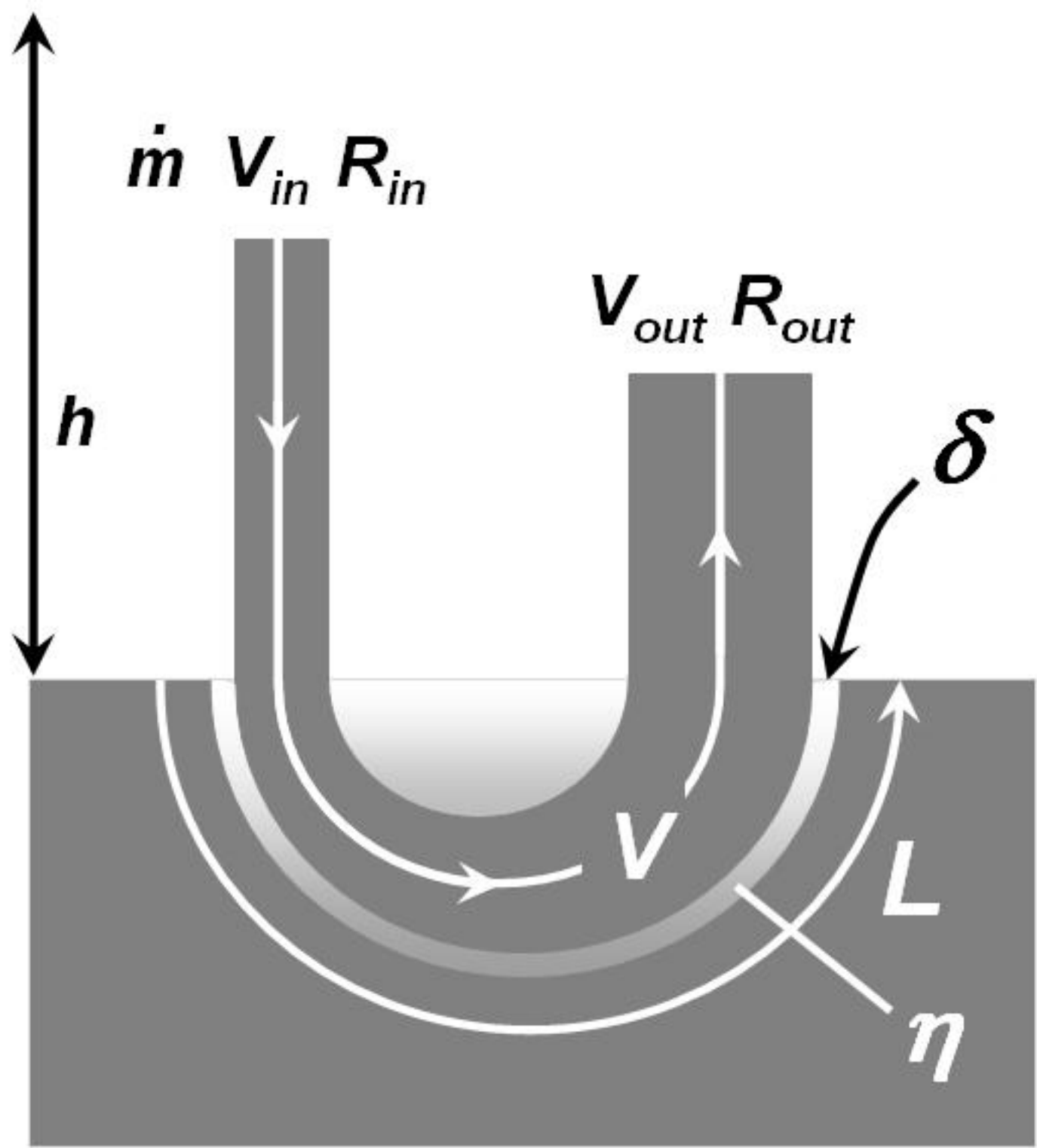

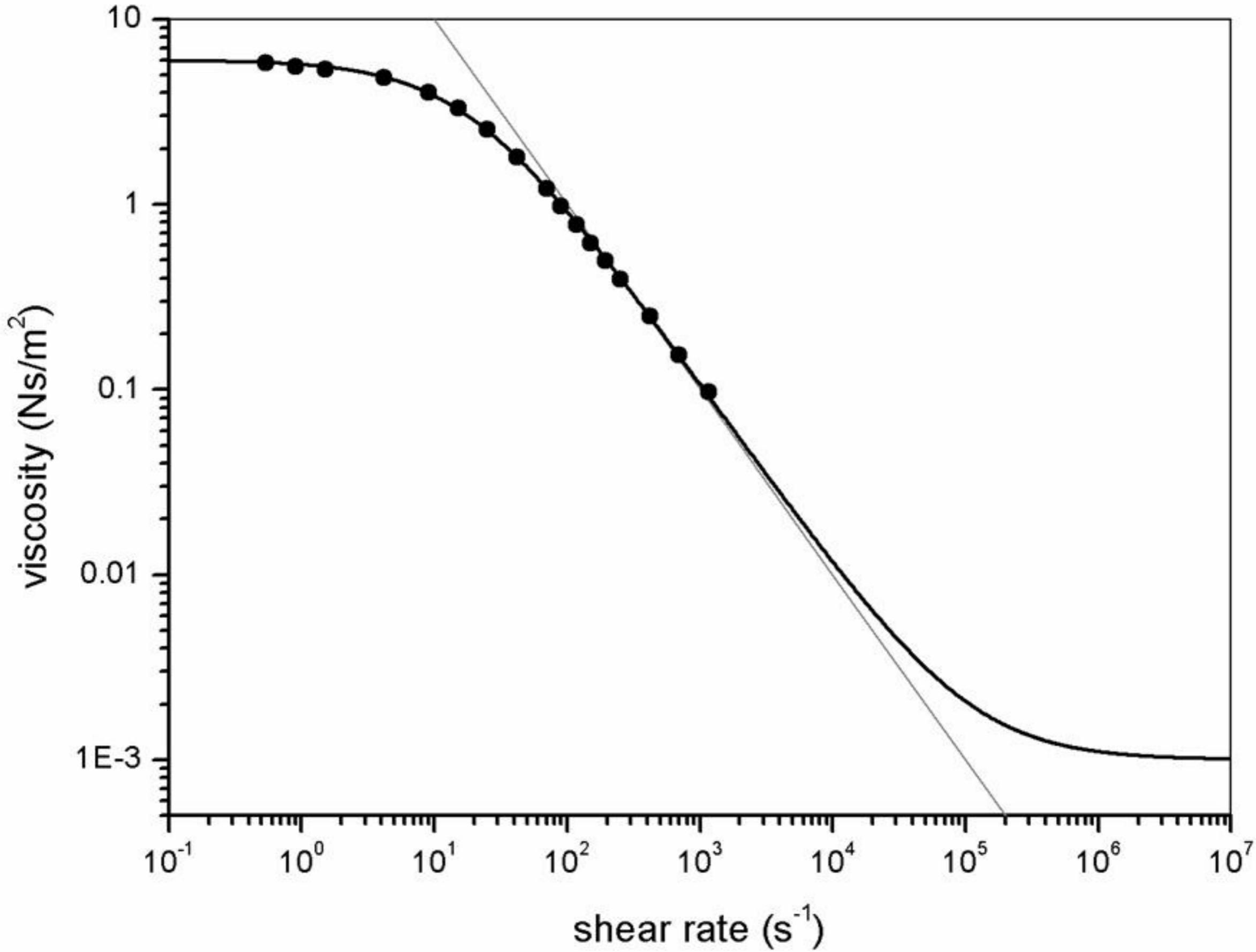

Fig. 3

Fig. 4

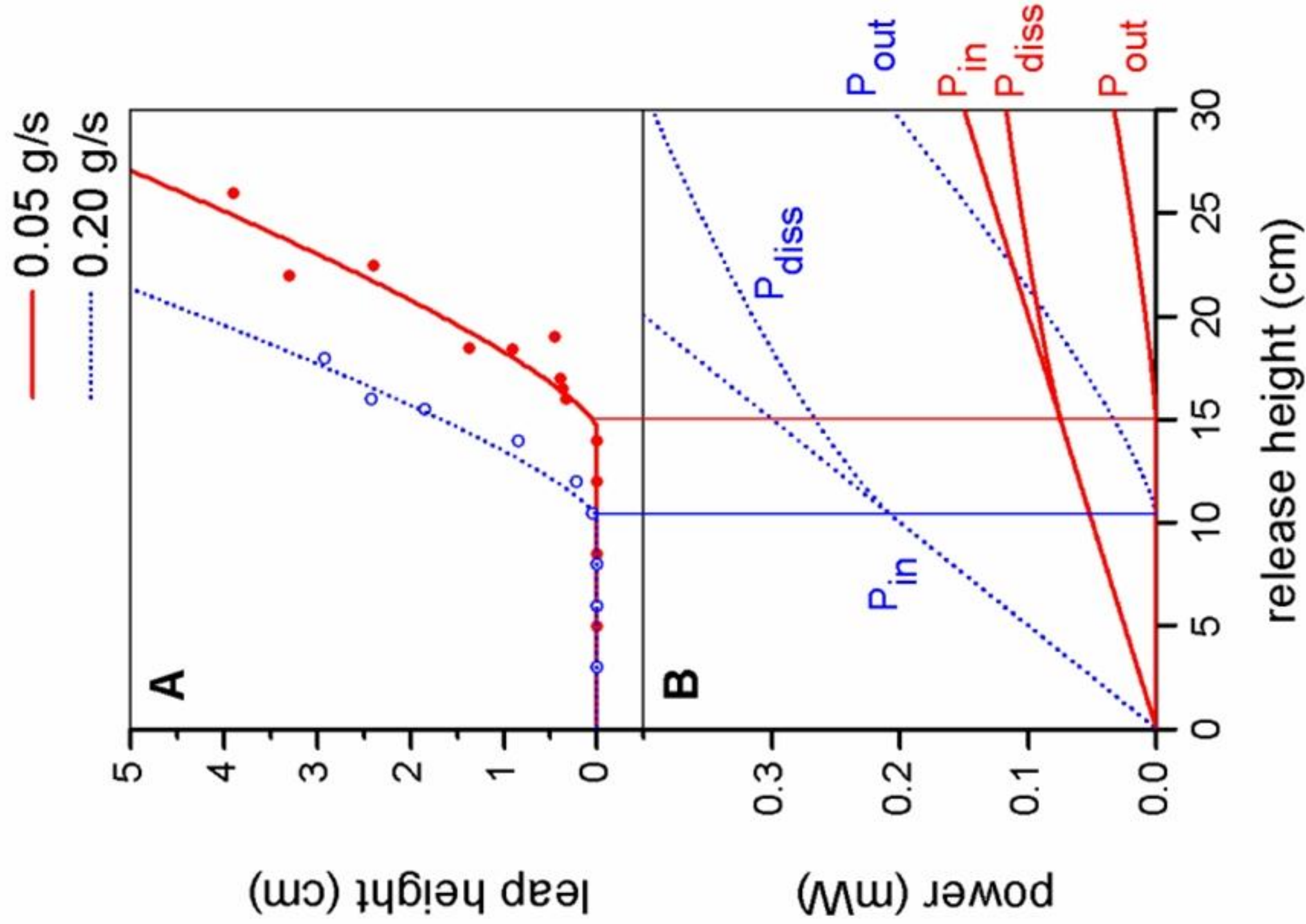

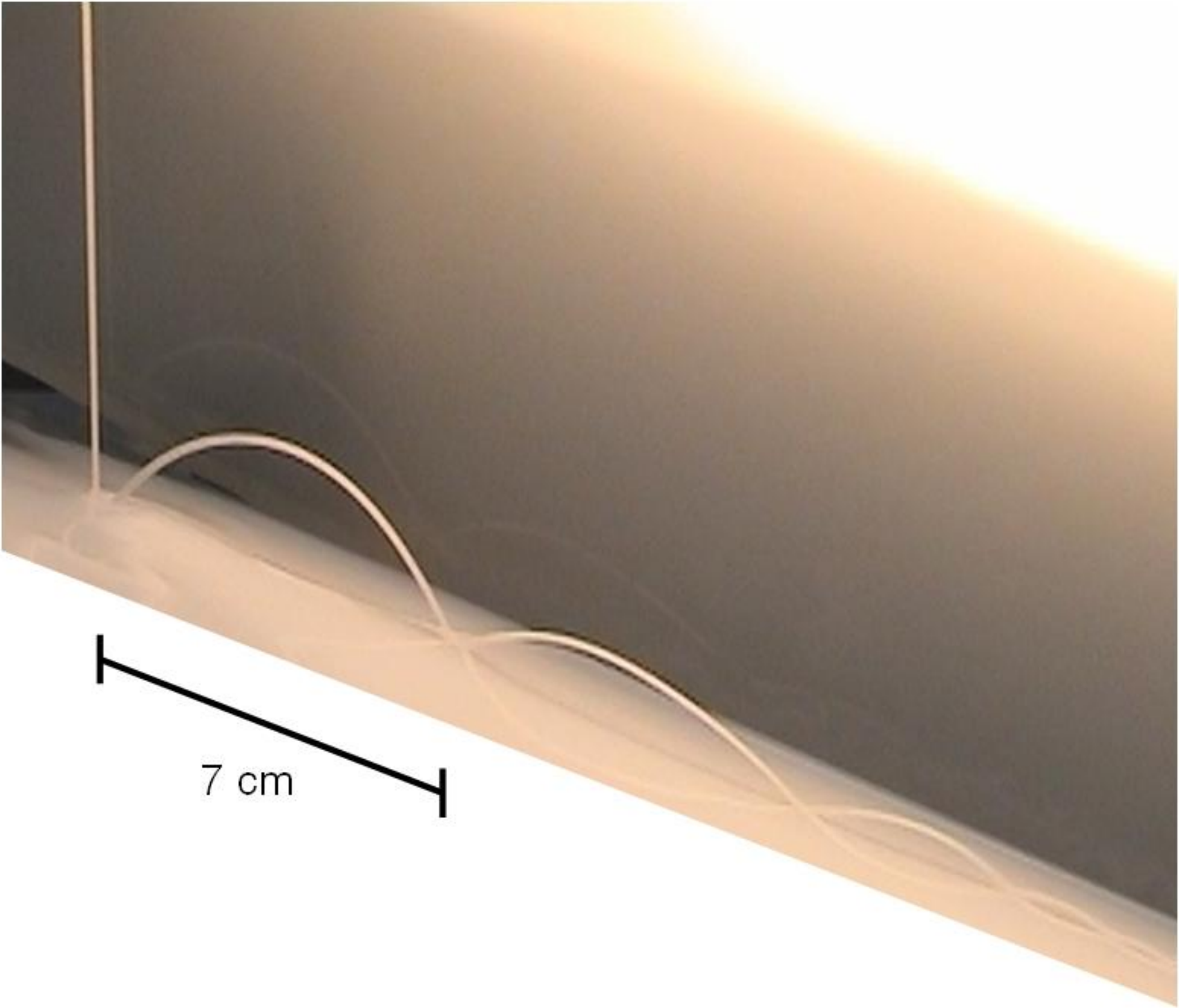

Fig. 5

7 cm

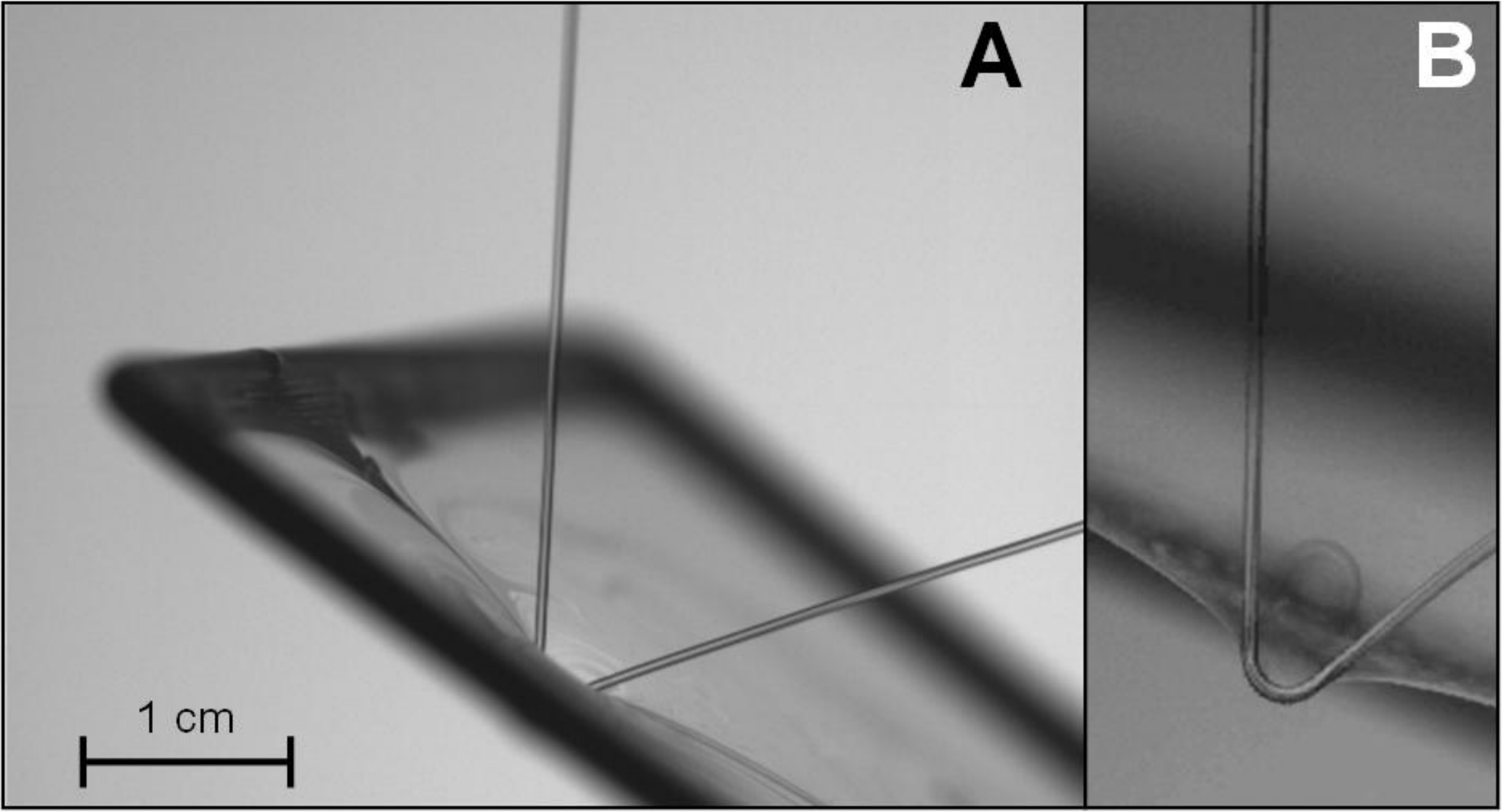

Fig. 6

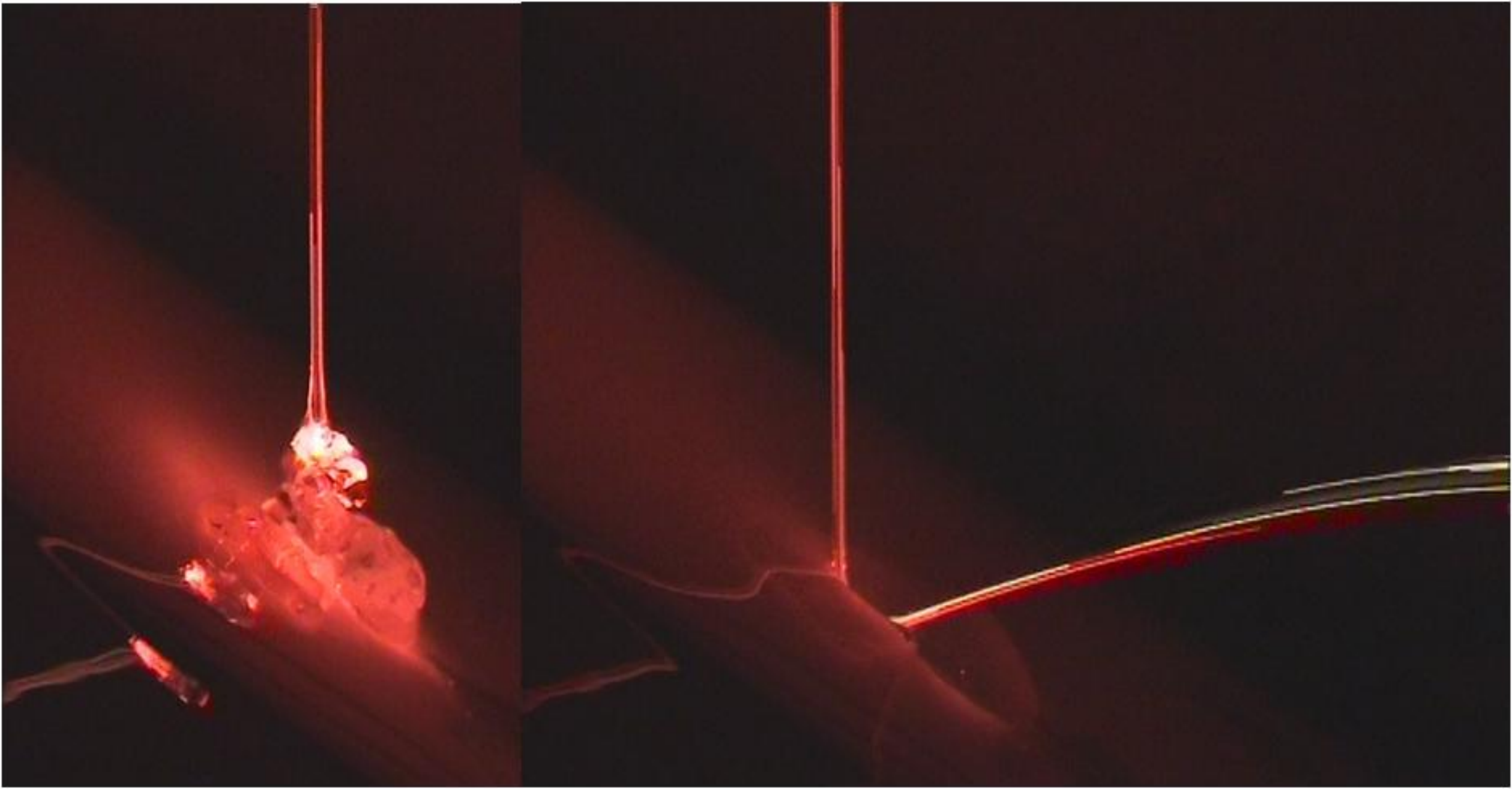

Fig. 7